\documentclass[pre,preprint]{revtex4}
\usepackage{psfig}

\begin{document}

\title{What thermodynamic features characterize good and bad folders? Results from a simplified off--lattice protein model}

\author{A. Amatori$^{1,2,3}$, J.Ferkinghoff-Borg$^4$, G. Tiana$^{1,2}$, and R. A. Broglia$^{1,2,3}$ }
\address{$^1$Department of Physics, University of Milano and} 
\address{$^2$INFN, sez. di Milano, via Celoria 16, 20133 Milano, Italy}
\address{$^3$The Niels Bohr Institute, Blegdamsvej 17, 2100 Copenhagen, Denmark}
\address{$^4$NORDITA, Blegdamsvej 17, 2100 Copenhagen, Denmark}

\date{\today}
\begin{abstract}
The thermodynamics of the small SH3 protein domain is studied by means of a simplified model where each bead--like amino acid interacts with the others through a contact potential controlled by a 20$\times$20 random matrix.  Good folding sequences, characterized by a low native energy, display three main thermodynamical phases, namely a coil--like phase, an unfolded globule and a folded phase (plus other two phases, namely frozen and random coil, populated only at extremes temperatures). Interestingly, the unfolded globule has some regions already structured. Poorly designed sequences, on the other hand, display a wide transition from the random coil to a frozen state. The comparison with the analytic theory of heteropolymers is discussed.
\end{abstract}

\maketitle

\section{Introduction}
Since Anfinsen first stated the {\em thermodynamic hypothesis} \cite{anfinsen} (that is, in a given environment, structural and functional features of proteins are fully encoded in their amino acids sequence) a consistent effort has been set in the study of the relationship between the amino acid sequence and its native structure and function. A significant part of this effort has been dedicated to the so called {\em inverse folding} problem, that is to the design of sequences which have a desired structure as unique, stable, kinetically accessible ground state (GS). The simplest approach to this problem is to search for the sequence which minimizes the energy of the system, keeping the native conformation and the ratio between the different kinds of amino acids (composition) fixed \cite{sh3}. At the basis of this approach lies the assumption that the free energy of most states of the system obey the principle of self-averaging, so that the total probability of the competing states is unaffected by the design. The property of self-averaging is also an element of the replica method, which complements the description given by the random energy model (REM) for heteropolymers \cite{sh1}.

A more efficient approach, which has given good results on lattice models, is to optimize either the Z-score \cite{david} or the approximated free energy of the system  \cite{deutsch, mich99}.  However, attempts to apply this idea to continuum hydrophobic-polar models has lead to results less satisfactory than expected \cite{mich98}.
Nevertheless, the {\em energy-minimization} approach has still the advantage of being simple to implement (especially in continuum space, where the wideness of the conformational space makes the calculation of the free energy non trivial) and it has proven successful in finding sequences which fold on the crystallographic structure of the SH3 domain within a dRMSD of $2.6 \AA$ \cite{andrea}.
There, as in other works \cite{colombo}, a major problem in achieving this goal has been the poor knowledge of the interaction among amino acids. A possible strategy to circumvent this limitation is based on the assumption that the ability of proteins to display a low-entropy equlibrium state at biological temperatures is a consequence of the heterogeneity of the interactions, together with the polymeric geometry of the system \cite{sh2}. 
Consequently, the {\em inverse folding} approach should work for any quenched random interaction, provided it is sufficiently heterogenous \cite{sh1} (and thus different from the simple hydrophobic-polar models).

\begin{figure}
\centerline{\psfig{file=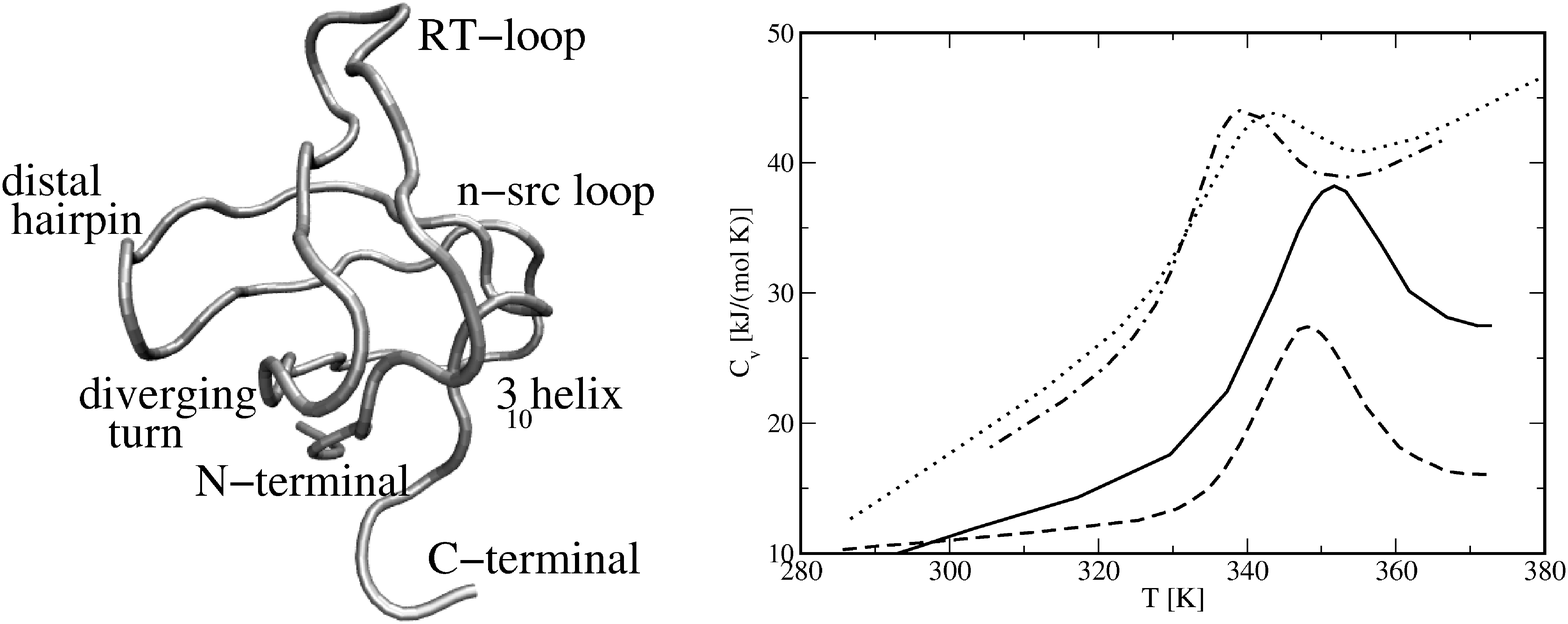,width=12cm}}
\caption{(Left) The native structure in a $C_\alpha$ representation of SRC SH3 as obtained by crystallographic experiments (pdb code 1FMK). On the picture is explicitely indicated the {\em RT-loop} (residues 8--19), the {\em Diverging turn} ($D_v$, residues 20--27), the {\em n-src loop} (residues 28--37), the {\em Distal hairpin} ($D_t$, residues 38--50), the helix $3_{10}$ (residues 51--54) and the sheet $\beta_{1-5}$ (residues 1--7/55--57). (Right) the specific heat of four different SH3-domain proteins, obtained from experiments, are shown (solid curve is Btk \protect\cite{cv1}, dashed curve is $\alpha$--spectrin \protect\cite{martinez98}, dotted curve is Abl, dash--dotted is Fyn \protect\cite{cv2}).}
\label{native}
\end{figure}

In the present work, we have focused our attention on SH3, a small (60 residues) $\beta$-like protein domain (cf Fig. \ref{native}, left panel) which has been widely investigated both experimentally \cite{martinez98, martinez99, riddle} and computationally \cite{guo, ding, hubner}.
Our model has proven successful in discriminating between good and bad folding sequences on the basis of only their native state energy. In particular, it has been possible to identify a threshold energy $E_{targ}^c$, such that sequences with native energy $E_{targ}<E_{targ}^c$ are good folders, while sequences with $E_{targ}>E_{targ}^c$ do not fold to the SH3 native structure and display low--energy conformations very diffent among themselves. 

Good folding sequences display a rather sharp transition between a globular unfolded state and the unique native conformation, as experimentally observed \cite{rose}.
Microcalorimetric experiments can explore temperatures which typically range from $0^o$ to $ 90^o C$. Experimental measures of the specific heat for four sequences folding to the SH3 domain are shown in the right panel of Fig. \ref{native}. All these proteins display a single peak in the specific heat at temperatures ranging from $\simeq 50^o $ to $\simeq 70^o C$. The thermodynamics of bad folders is more difficult to study due to their tendency to clump into insoluble aggregates.

In the present work we employ efficient sampling algorithms \cite{jesper} to study the thermodynamics of good and bad sequences designed on the SH3 fold and compare them to random sequences. The main issues we want to investigate are the nature of the equilibrium states that the protein populates, as well as the thermodynamics and structural properties of these states. Furthermore, we study the extent to which these properties can be described by the standard theory of heteropolymers \cite{pande,khokhlov}.

\section{The model} \label{sect_2}

The model we use has been described in ref. \cite{andrea}. It is a reduced off-lattice single-bead model where the amino acids are represented by spherical beads centered around the C$^\alpha$-atom connected by an inextensible chain. The energy potential is the sum of pair interactions via a square well function where each $\sigma$-{\em th} amino acid type is characterized by a specific value of the hard core radius $R^{HC}(\sigma)$ and of the interaction strenghts $B(\sigma,\pi)$. The matrix $B(\rho,\pi)$ is generated according to a Gaussian distribuition (the mean is $B_0=0.23$ and the standard deviation is $\sigma_B=0.53$, in arbitrary units) and sequences are designed making use of a Monte Carlo simulation at various temperatures in the space of sequences. That is, switching two amino acids at random and then accepting or rejecting the change according to the {\em Metropolis} algorithm \cite{metropolis}.

Another important ingredient of the model is a constrain on the total number of contacts each residue is allowed to build. This constrain has been introduced because single-bead models oversimplify the geometry of the residues and thus give rise to unphysical conformations where residues build more contacts than their real geometry would allow. Therefore a maximum number of contacts $n_{max}(\sigma)$ has been assigned to the 20 amino acids.
The thermodynamic sampling has been performed making use of a generalized weights algorithm \cite{jesper}. 

The order parameters analyzed are the Radius of Gyration ($R_g$), the RMSD ({\em Root Mean Square Deviation}) and the dRMSD \footnote{dRMSD is defined as the root of the mean square difference between the inter--residue distance in the given conformation and in the native state, calculated over all pairs of residues.} (distance root mean square deviation), which performs well in discriminating different states of the system. In all our simulations RMSD and dRMSD resulted to be highly correlated. Consequently we will only refer to dRMSD.
When dRMSD is used to calculate geometrical differences between a given structure and the native src-SH3 we will use the simbol $d_N$. When used to calculate differences between any couple of structures, it will be called $d_S$. 

For six particular secondary structures of src-SH3 (that is the {\em RT-loop} (residues 8--19, cf. Fig. \ref{native}), the {\em Diverging turn} ($D_v$, residues 20--27), the {\em n-src loop} (residues 28--37), the {\em Distal hairpin} ($D_t$, residues 38--50), the helix $3_{10}$ (residues 51--54) and the sheet $\beta_{1-5}$ (residues 1--7/55--57) ) as well as for their relative conformations, we define a {\em ``structure content''} $q$ as the average fraction of native contacts within each structure.

In what follows, we study the thermodynamic behaviour of 9 different sequences summarized in Table \ref{tab_seq} and generated according to the $E_{targ}$ criteria. 

\section{Thermodynamics of good folders}

\begin{figure}
\centerline{\psfig{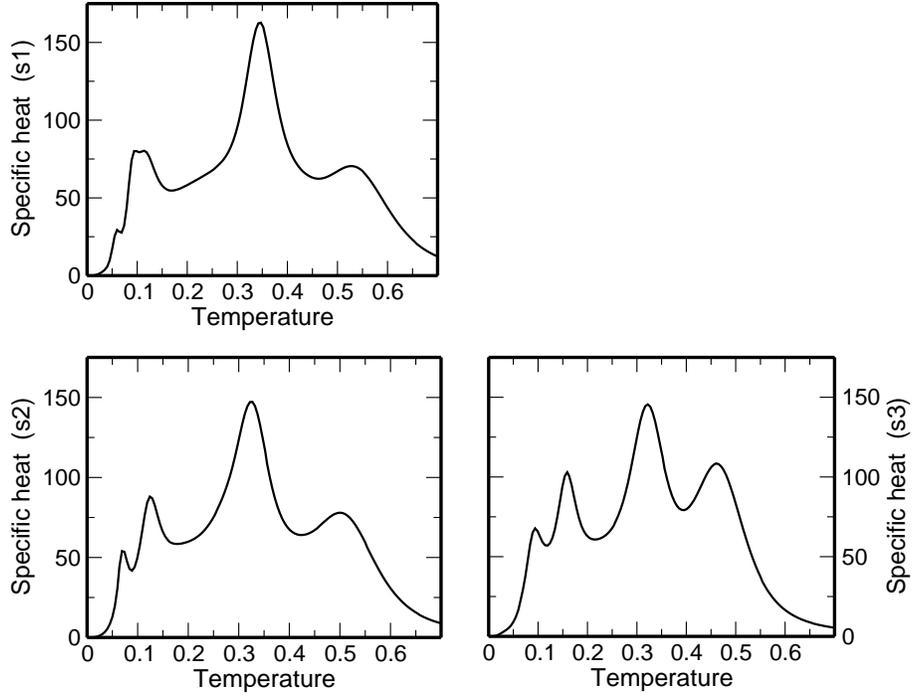}}
\caption{The specific heat $C_p(T)$ of the three good folders for sequence $s_1$ (top), sequence $s_2$ (bottom left) and sequence $s_3$ (bottom right)}
\label{calGood}
\end{figure}

\begin{figure}
\centerline{\psfig{file=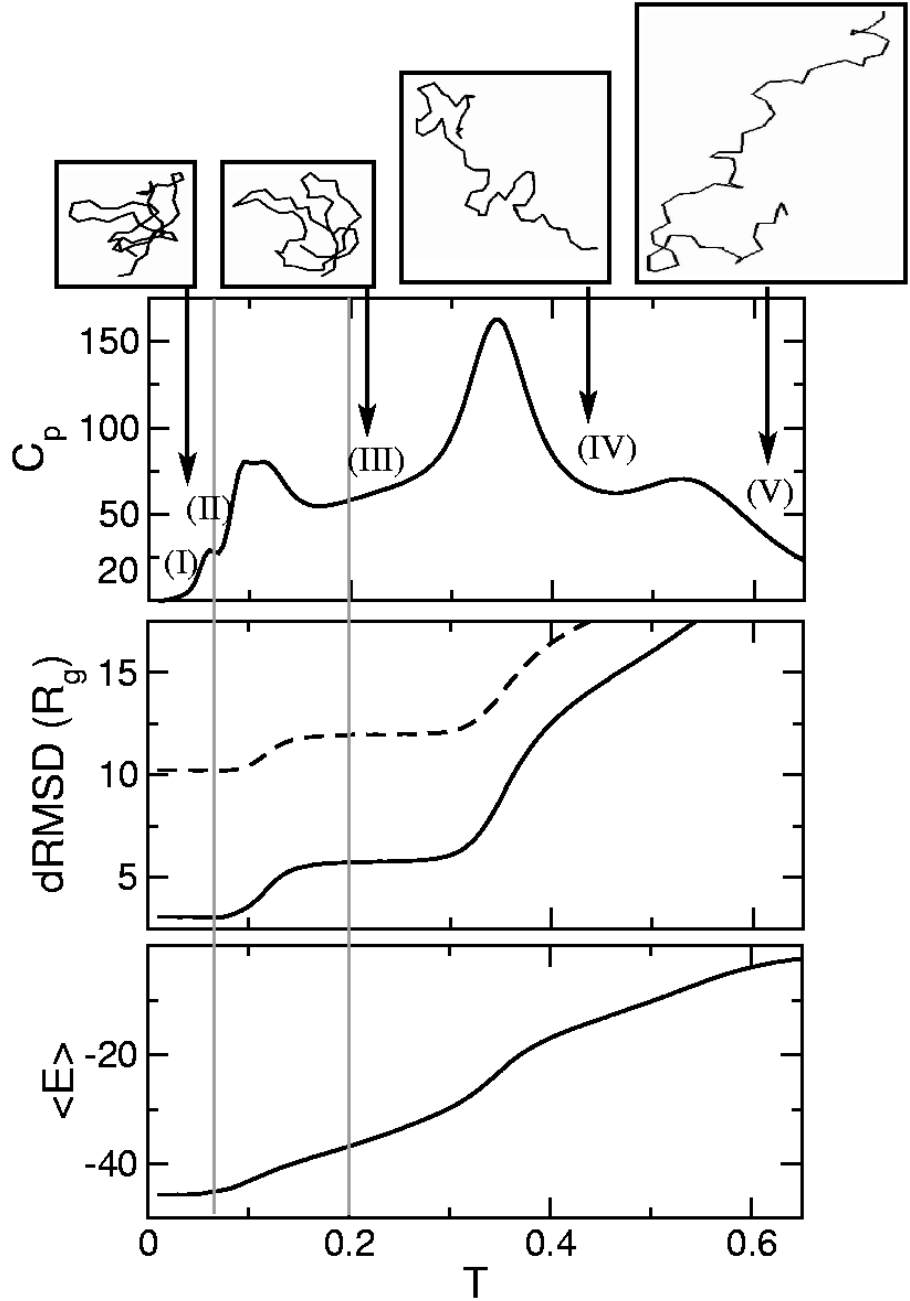,width=12cm}}
\caption{The specific heat $C_p(T)$ (upper panel), average $d_N$ expressed in $\AA$ (middle panel, solid curve), average radius of gyration $<R_g>$ (middle panel, dashed curve) and average energy $<E>$ (lower panel) for sequence $s_1$. The interval of temperatures defined by gray perpendicular lines marks the region of biological relevance. The four pictures on the top show typical conformation of the system in each state (the first picture represents both state (I) and state (II), as conformational differences are neglegible on this scale).}
\label{s1}
\end{figure}

Sequences $s_1$, $s_2$ and $s_3$ (cf. Table \ref{tab_seq}) display protein--like properties \cite{andrea}, having a unique and stable native state corresponding to the SH3 conformation shown in Fig. \ref{native} and being able to reach it in a short time. 
The conformational specific heat $C_p(T)$ of these sequences, calculated with the present model, is displayed in Fig. \ref{calGood}. An interesting feature is that the three sequences, although having less than 10\% identity, display very similar specific heat. $C_p$ is in all cases characterized by four peaks, which mark the transition between different states. Because of these similarities, in the following we refer to the behaviour of sequence $s_1$, as a template for all good folders, unless otherwise mentioned.
To identify the features of the thermodynamical states of this sequence, we have plotted in Fig. \ref{s1} the averages of $d_N$, $R_g$ and of the energy $E$ as functions of temperature. At high temperatures ($T>0.6$, the corresponding thermodynamical state being marked as (V) ) the chain behaves as a {\em random coil} with an average energy only slightly below zero.
In this state the protein has very few ($< 5$) contacts, no detectable secondary structures, a mean gyration radius $\bar{R}_g \simeq 22 \AA$ and a $d_N$ between $18\AA$ and $25\AA$. That is, it does not have anything in common with the native conformation.

Decreasing the temperature, the system shows a low, wide, peak in $C_p(T)$. The state lying beyond the peak (state (IV) in Fig. \ref{s1}) is still extended, having a mean radius of gyration of $18\AA$. The associated conformations are overall dissimilar from the native one, with $\bar{d}_N =15\pm 4\AA$ and $\bar{d}_S =9\pm 2\AA$. The equilibrium distribution of $d_S$ at $T=0.5$ is shown with a dashed curve in the upper panel of Fig. \ref{p_drmsd} and indicates a wide structural heterogeneity. Nonetheless, state (IV) has a sizable content of structured {\em Distal hairpin} and {\em RT-loop} ($q$ between 0.25 and 0.56), while the {\em n-src} loop, the {\em Diverging turn}, the helix $3_{10}$ and the sheet $\beta_{1-5}$ are essentially absent (cf. Table \ref{tab_TS_G}). These conformations are characterized mainly by local bonds, although there are few non-local native contacts of residues 2, 3 and 4 with residues 24, 25 and 26, which give rise to the RT-loop.

At a temperature $T \simeq 0.34$, the system undergoes a marked decrease in the energy and a sharp compaction ($\bar{R}_g$ decreases from $18\AA$ to $12\AA$).
The state beyond this {\em coil--globule transition} (state (III) of Fig. \ref{s1}) is associated with conformations displaying an average $d_N =6\AA$, still dissimilar from the native conformation, thus qualifying as unfolded state. This transition is underlined by a rather sharp peak in $C_p$, consistently with a first order coil--({\em ordered}-)globule transition, as predicted by the theory of non-random heteropolymers \cite{pande}. The distribution of $d_S$ for state (III) ($T \simeq 0.2$) is shown in Fig. \ref{p_drmsd} (dotted curve). Although the major peak is still centered at $d_S\simeq 8\AA$, indicating the structural heterogeneity typical of the unfolded state, a small peak emerges at $d_S=4\AA$, which indicates a small presence of specific conformations.
On the other hand in this region the specific heat is well above zero, and the energy displays values in the range $E\simeq -29$ to $E\simeq -39$. Interestingly, this energy interval is not accompanied by any major structural changes, as indicated by the approximately constant values of $R_G$ and $d_N$.

Analyzing the conformations and the map of contacts of state (III), we observe that it is characterized by the RT--loop and the $D_t$ essentially fully formed (with probability 0.83 and 1.0, respectively), and, consequently, the sheet between strands $\beta_3$ (36--41) and $\beta_4$ (47--51). Also the Diverging turn $D_v$ is well structured at this stage ($q=0.75$), while the two terminals get together and give rise to a shorter sheet (which is generally not fully formed at this point; $q=0.55$) between strands $\beta_1$ and $\beta_5$. The poor presence of n-src loop ($q=0.40$) causes the sheet $\beta_2$ (24--28) - $\beta_3$ not to be  formed. Within this context one can see from Table \ref{table_new} as well as Fig. \ref{fig_new} that the contacts between $D_t$ and $D_v$ are not formed.  On the other hand, one finds that the contacts between the {\it RT--loop} and the distal hairpin are formed with high probability (0.86), contacts which were not formed in IV.
Summing up, the presence of the ensemble described above, dominated by unfolded globular conformations displaying structured fragments is a thermodynamic hallmark of good folders.
 
Going back to Fig. \ref{s1}, a third peak is found at temperature $T=0.10$ and marks a transition to a state with average $d_N=3.0\AA$ and RMSD$=4.6\pm0.8\AA$, which can be regarded as the native state (cf. ref.\cite{andrea}).  This transition gives rise to the formation of the n-src loop, which is associated with a large entropy loss. Moreover, at this temperature the system undergoes the formation of the helix (with 75\% probability and an increase in the content of sheet $\beta_{1-5}$ ($q=0.85$, cf. Table \ref{tab_TS_G}) which, nontheless, is still able to fluctuate. The average radius of gyration of state (II) is $\bar{R}_g=10\AA$, which is the same as that of the crystallographic native conformation and is only 20\% more compact than state (III). The distribution of $d_S$ is now peaked around $2.5\AA$ (cf. Fig. \ref{p_drmsd}), in accordance with the features of uniqueness of the native state. These data suggest that the peak at $T=0.10$ is the one observed in calorimetry experiments (see Fig. \ref{native}), associated with the {\em folding transition} (in Fig. \ref{s1} the hypothetical experimental window is indicated with a gray frame, assuming the $T_{fold}$ of SH3 to be $333 K$ \footnote{ Guo {\em et al.} \cite{guo} estimate the folding temperature for src-SH3 to be within 323 K and 343 K}). Note that the other peaks of Fig. \ref{calGood} correspond to much larger temperatures, and consequently can hardly be observed under normal experimental conditions.

At temperature $T=0.06$ the specific heat of sequence $s_1$ has a last, small peak and then sharply drops to zero, in correspondence with the {\em freezing} of the system into its ground state. The contact map of state (I) indicates that the only essential structural difference from the previous state is a tightened sheet $\beta_{1-5}$ ($q=0.90$), that freezes the last degrees of freedom of the system (cf. Table \ref{tab_TS_G}).
 
\section{Thermodynamics of bad and random sequences}

For comparison with the good folding sequences, we next analyze the specific heat of three randomly--generated sequences (sequences $s_7$, $s_8$ and $s_9$) and of three bad folders ($s_4$, $s_5$ and $s_6$), that is sequences designed to have, on the SH3 native conformation, an energy too high to fold (i.e. larger than $E_c$ \cite{andrea}). The plots of the specific heat are displayed in Figs. \ref{calRand} and \ref{calBad}, respectively.

In all these cases the pattern common to good folders is lost and the specific heat have a more sequence--dependent shape. The shape of $C_p$ for bad folders is characterized by a large shoulder which involves all temperatures up to $T=0.6$, with peaks superimposed in a disordered fashion. On the other hand, random sequences display a more compact $C_p$, being significantly different from zero only in the region between $T=0.05$ and $T=0.40$.
The peaks in the specific heat of bad and random sequences are tightly connected with the variation of the radius of gyration of the protein. In Fig. \ref{gyradCompa} the average radius of gyration $\bar{R}_g$ for a folding sequence ($s_1$, solid curve), for a bad sequence ($s_4$, dashed curve) and for a random sequence ($s_7$, open circles) are shown. 

\subsection{Random sequences}

The radius of gyration of the random sequence displays a wide sigmoidal shape which spans the region of the major peak in the specific heat. Consequently, this wide peak is associated with a broad transition from a coil ($\bar{R}_g\simeq 20\AA$) to a compact globular phase ($\bar{R}_g\simeq 10\AA$). The range of temperatures which can be interpreted as biologically relevant (cf. Fig. \ref{s1}) partially overlaps the peak in $C_p$. 
According to Flory's model of homopolymer collapse \cite{flory,witelski}, the volumetric interaction free energy takes the form:
$$
F_{vol}(T,\phi)\simeq {\cal N} T\left( 1-\chi\phi + \frac{1-\phi}{\phi}\log(1-\phi)\right),
$$
where $\phi= \nu {\cal N}/V$ is the average polymer volume fraction, ${\cal N}$ the total number of monomers in the chain, $\nu$ is the excluded volume of each monomer and $V$ is the average volume occupied by the chain, $V\simeq \left(\frac{5}{3}\right)^{3/2}\frac{4\pi}{3}R_g^3$ \footnote{The radius of a maximally compact spheric protein ($\phi=1$) is $R=\left(\frac{3}{4\pi} {\cal N}\nu\right)^{1/3}$, whereas the volume, $V_g$, obtained from the radius of gyration is $V_g= \frac{4\pi}{3} R_g^3 = \left(\frac{5}{3}\right)^{-3/2} \frac{4\pi}{3} R^3$. In order to find the correct result in the limit $\phi\rightarrow 1$, we therefore calculate the average volume as $V=\left(\frac{5}{3}\right)^{3/2}V_g$}. From the maximally dense globule obtained from simulations, $R_g\simeq 9.8 \AA$, we obtain $\nu\simeq 141~\AA^3$ as the average excluded volume. The Flory-Huggins constant, $\chi$, can be estimated from the number of contacts in the dense globule, $N_C$, by noting that ${\cal N}\phi$ is the number of binary collisions in the mean-field picture, hence $E=-{\cal N} T\chi\phi$ is the total interaction energy. In the high density limit ($\phi\rightarrow 1$), one has $E=-N_CB'$, where $B'$ is the effective interaction energy between monomer pairs. Thus, $\chi\simeq - \frac{N_C}{\cal N}\frac{B'}{T}=-\frac{z}{2}\frac{B'}{T}$, 
and $z$ is the coordination number for the (dense) globule. In the case of a homopolymer, $B'=B_0$,  where $B_0$ is simply the monomor-monomer interaction strength.  In the heteropolymeric case, the effective interaction is modified according to $B'=B_0-\sigma_B^2/2T$ \cite{pande}, where $B_0$ now denotes the average of the interaction matrix, $B_0=\sum_{\alpha\beta} p_\alpha p_\beta B_{\alpha\beta}$, and $\sigma_B^2 = \sum_{\alpha\beta} p_\alpha p_\beta (B_{\alpha\beta}-B_0)^2$ is the variance. Here, the summation, $\sum_{\alpha\beta}$, runs over the different monomer types, $\alpha$ and $\beta$, and $p_\alpha$ is the frequency of occurence of monomor type
$\alpha$. In our case, $B_0=0.23$,  $\sigma_B= 0.53$ and $z\simeq 2.2$. Combining the various expressions above and expanding the total interaction free energy for a random heteropolymer to third order in the density we obtain 
$$
F_{vol}(T,\phi)= {\cal N} T \left[-\frac{z}{2}\left(\frac{\sigma_B^2}{2T^2}-\frac{B_0}{T}\right)\phi+\frac{1}{2}\phi+1/6\phi^2+{\cal O}(\phi^3)\right],
$$
with the corresponding second and third virial coefficients $b(T)=\frac{\nu}{2T}\left( T-z\frac{\sigma_B^2}{2T}+z B_0\right)$ 
and $c=\nu^2/6$, respectively. The theory predicts a second-order {\em coil--globule transition} at temperature  $\theta$ where $b(\theta)=0$. 
In the present case, $\theta\simeq 0.36$, which is in the high end of the transition region. However, according to 
the standard Lifshitz theory of the coil-globule transition \cite{khokhlov} there is an entropy cost associated with
the surface formation of the globule, because the chain sections on the surface layer necessarily have the form of loops.
Since this entropy cost scales with the system size as ${\cal N}^{2/3}$, the transition temperature, $T_{cg}$, may be 
shifted to a value somewhat below the thermodynamic $\theta$-point for small systems. By balancing
the  energy gain from the coil collapse with the entropy loss of the surface formation the theory predicts the
relative shift of the transition temperature, $\tau_{cg}\equiv \frac{T_{cg}-\theta}{\theta}$, to be \cite{khokhlov}
\begin{equation}
\tau_{cg}\simeq -2.7 a^{3/2}c^{1/4}b_\theta^{-1}{\cal N}^{-1/2},
\label{trans}
\end{equation}
where $a$ is the Kuhn length and $b_\theta$ is defined from the Taylor expansion of $b(\tau)$ around the theta point, $b(\tau)=b_\theta\tau+{\cal O}(\tau^2)$ and $\tau=\frac{T-\theta}{\theta}$. For the present model, we obtain
$b_\theta=\nu\left(\frac{1}{2}+\frac{z\sigma_B^2}{4\theta^2}\right)$. To determine $a$, we assume for simplicity that the chain behaves as a gaussian coil at the point where $\frac{dR_g}{dT}$ has its maximum. From the relation $R_g^2=\frac{1}{6}al{\cal N}$, where $l\simeq 3.8 \AA$ is the distance between consecutive monomers, one obtains $a\simeq 7.6 \AA$, corresponding to a moderately flexible chain, $v/a^3\simeq 0.32$. Inserting the expression for $b_\theta$ and $c$ in Eq. (\ref{trans}) gives
$$
\tau_{cg}\simeq -2.7\cdot 6^{-1/4}\left[\frac{1}{2}+\frac{z}{4}\left(\frac{\sigma_B}{\theta}\right)^2\right]^{-1}
\left(\frac{a^3}{{\cal N}\nu}\right)^{1/2}\simeq -0.23 ,
$$
corresponding to $T_{cg}\simeq 0.27$, which is in excellent agreement with the observed midpoint of the transition, cf. Fig. \ref{calRand}. The coil--globule transition happens, coherently with the theory of random heteropolymers \cite{pande}, at approximately the same temperature ($T \simeq 0.25$) for all the random folders. 

From the lack of any plateau in $\bar{R}_g$ (see Fig. \ref{gyradCompa}), we see that the system is not populating any well-defined globular state in the transition from the coil to the frozen ground state. In other words, differently from good folders, the freezing transition is not immediately distinguishable from the coil--globule transition region. 

Making use of the common picture of freezing \cite{pande} as the process which brings random heteropolymers from the sea of random globular configurations into their ground state basin, we calculate the thermodynamical average of $d_S$ as function of temperarature.     
The plot of $\bar{d}_S(T)$ (cf. Fig. \ref{dS_rand} (a)) shows indeed a marked transition for all the sequences from values $\bar{d}_S\approx 1.5 \AA$ (corresponding to mutually similar structures) to $\bar{d}_S\approx 4.2 \AA$ 
(corresponding to compact structures with no similarity). The average temperature of this {\em freezing transition} is $T_{freeze} = 0.1$ (cf. Fig. \ref{dS_rand} (a)) and coincides with a sharp decrease of $C_p(T)$.
In accordance with REM \cite{pande}, which predicts the freezing transition to be of second order, we may regard this decrease of $C_p$ as a signature of a second order phase transition in our finite system.

To investigate the roughness of the energy landscape, we calculate the value of $d_S$ between low energy states, chosen within the 0.15-fractile of the energy distribution at $T = T_{freeze}$ (i.e. $P_T( E < E_{15\%})= 0.15$, at $T=T_{freeze}$).
As Fig. \ref{dS_rand} (b) shows, the set of these low energy ($E < -39$) states for random sequences is very heterogenous ($\bar{d}_S = 4.4 \AA$, and null probability for $d_S < 2.7 \AA$). On the other hand, kinetic simulations below the calculated freezing temperature ($T < 0.1$), initialized in any one of these low energy conformations, visit states with a pairwise distance $d_S < 2.5 \AA$ (cf. Fig. \ref{dS_rand} (c)). This shows that conformations with $d_S > 2.7 \AA$ are typically separated by consistent energy barriers which make them kinetically inaccessible to each other even at temperatures where the specific heat is well above zero. This is fully consistent with the thermodynamics of random heteropolymers \cite{sh1}, which predicts a free energy landscape at low temperature with several wells, each well containing conformations mutually similar, and different wells containing conformations with little similarity. The low--temperature state of the random sequence is then, according to the language of ref. \cite{sh1}, a {\em frozen state}.

The theory of heteropolymers predicts the freezing temperature of large globules to be $T_{freeze}=\frac{\sigma_B}{2\sqrt{\Delta s}}$ \cite{sh1}, where $\Delta s$ is the entropy per contact lost in the freezing. Using as an estimate $\Delta s\simeq \log(a^3/\nu)\simeq 1.14$ \cite{pande}, one obtains $T^{theor}_{freeze}\simeq 0.25$, which is significantly higher than the value
estimated from the simulations.  Interestingly, the crude approximation $\Delta s\simeq \log(a^3/\nu)$ is 
surprisingly close to the observed value $\Delta s= 1.01$ \footnote{The use of $\sigma_B$ as the standard deviation per contact of the energy probability distribution assumes each contact to be an independent random variable, which can only be expected to hold in the flexible chain limit $\nu/a^3 \rightarrow 1$.
Assuming the actual number of uncorrelated contacts ($\tilde N_c$) to depend on the Kuhn length as $\tilde N_c=N_cl/a$,
implies that the standard deviation of the energy probability distribution is reduced according to $\sigma_B\rightarrow \sigma_B\sqrt{l/a}$. In fact, this simple rescaling is in good agreement with the observed standard deviation of the energy probability distribution around $T=T_{freeze}$ (data not shown) and reduces consistently the predicted freezing temperature ($T^{rescaled}_{freeze}\simeq 0.18$). Nonetheless the freezing temperature obtained from this rescaling is still significantly higher than the observed $T_{freeze}$, indicating residual correlations between different states.}.
   
Importantly, the observed $T_{freeze}$ is also low compared with the critical design temperature of our model ($T^{cr}_{design} \simeq 0.15$), in marked contradiction with the prediction of REM, where $T^{cr}_{design}=T_{freeze}$. The failure of the equations $T_{freeze}=\frac{\sigma_B}{2\sqrt{\Delta s}}$ and  $T^{cr}_{design}=T_{freeze}$ shows that, in our system, low energy states are not uncorrelated, implying that the principle of self-averaging does not in general apply.

From the structural point of view, random sequence display surprising features. In Table \ref{tab_TS_R} the features of the random sequences are reported for their three thermodynamically relevant states.
As far as the formation of secondary structures is concerned, random sequences are obviously less effective than good folders. 
In particular, the helix and the sheet $\beta_{1-5}$ are practically absent at all temperatures in random folders.
Nonetheless, in most of them, the presence of the other peculiar structures (i.e. the {\em RT-loop}, the {\em Diverging turn}, the {\em n-src} loop and the {\em Distal hairpin}) is non-neglegible in low energy states.
This highlights the fact that these structures, primarely based on local and mid-range bonds, pay a lower entropy and hence require less design accuracy to be formed. In other words, there is a structural imprint based only on the length of the chain which favours turns in specific places, and the evolutionary optimization of good sequences can take advantage of such imprint to imcrease its stability.

\subsection{Bad sequences}

While the rationale for the specific heat of good folders resides in their common folding properties and that for the random sequences in their average properties, bad folders seem to have lost both of them. Their specific heat is featured neither by the specificity of their sequence, nor by the averaging of the contribution of uncorrelated residues. Even the coil--globule transition is overwhelmed by other effects, and is not easily detectable in the plot of $C_p$ (cf. Fig. \ref{calBad}). 

The $\bar{R}_g(T)$ plots of these sequences show a rather smooth decrease (as temperature decreases) from random coil values above $20 \AA$ to a ground state value $\simeq 10.5 \AA$ (see Fig. \ref{gyradCompa} where the $\bar{R}_g(T)$ plot of sequence $s_5$ is shown). This implies that, similar to random sequences, bad folders do not have the globular unfolded state (III) typical of good sequences. A qualitatively similar behaviour is also found for $d_N$ plots (data not shown).

Unlike random sequences, their low energy states do not populate a multitude of minima. 
In fact, thermodynamical samplings of the kind used for good and random sequences, starting from a random conformation and lasting for $50 \times 10^{9}$ steps, find a non-native (${\overline d}_N= 5.0\AA$) but still structurally homogeneous (${\overline d}_S=3.1\AA$) basin, the bottom of which displays an energy $E \simeq -44$. Nonetheless, fixed temperature Monte Carlo simulations at $T=0.1$ starting from the crystallographic conformation of SH3, display {\em another} homogenous ($\bar{d}_S= 2.4 \AA$ and maximum of $d_S = 4.1 \AA$) basin with practically the same energy ($E_{min} = -44.5$) but strongly dissimilar from the first one (crossed $\bar{d}_S = 4.7 \AA$ with minimum value of  $d_S = 3.6 \AA$). The native conformation belongs to this second basin (whose $\bar{d}_N = 3.2 \AA$), i.e. the native basin is a local minimum of the free energy, at least for temperatures up to $0.1$ (where the $C_p$ is still consistently above zero).
But while performing the simulation at $T = 0.3$ the system leaves the native state almost immediately, reaches states belonging to the non-native low-energy state and is never able to return to the native basin within the $5 \times 10^{9}$ steps of the MC sampling.

Our generalized-weight sampling algorithm does not implement an order parameter capable of differentiating between the two basins (which span the same range of energies) and thus is not able to provide their respective free energies. On the other hand the above results show that, at not--too--low temperatures, there is not a large barrier separating the native from the non-native basin, hence the entropy of the non-native state must be markedly larger than that of the native, while their energies are similar.
                                                                                                                             
Unlike good folders (which compensate the low entropy of the native state by optimizing the sequence in such a way that {\em E(Native basin) $<$ E(Random globule)} ), the picture which emerges here is that the native state of bad folders is still entropically disfavoured, but does not have an energetic advantage ({\em E(Native basin) $\simeq$ E(Random globule)} ) to counterbalance the entropy of competing states. 

The alternative ground state of bad folders does not satisfy the conditions of kinetical accessibility and thermodynamical stability required to a protein's native state. 
In fact, kinetic simulations at fixed temperatures ranging between $T= 0.1$ and $T= 0.2$, reach this state only about four out of ten times, while the other six times get trapped into local energy minima (rough energy landscape). 
Furthermore, the plot of $dRMSD_{GS}(T)$ (Fig. \ref{dGScompa}) shows that this ground state is not even thermodynamically stable. Indeed $dRMSD_{GS}$ has a smooth increase with the temperature, i.e. there is no energetic barrier segregating the ground state from other higher energy states.

Of course, these results are not able to exclude the existence of other basins, although they have not been observed in very long simulations for each of the three bad sequences.

Table \ref{tab_TS_B} summarizes the features of the bad folders in the three themodynamically relevant states. 
The analysis of secondary structures formation confirms what stated when random and good sequences have been compared.
Indeed the {\em RT-loop}, the {\em Distal loop} and, to a less extent, the {\em n-src} loop and the {\em Diverging turn}  have a non-neglegible presence in both the frozen and the globular state of all bad sequences.
On the other hand the helix and the sheet $\beta_{1-5}$ have a comparatively low presence in all sequences at any temperature.
Remarkably, the values of $q$ in low energy states of bad folders reflect the hierarchy of the same structures in good folders (cf. Table \ref{tab_TS_G}).
Namely, the {\em RT-loop} and  the {\em Distal loop}, with an average value of $q$ at the frozen state of bad folders of respectively 0.59 and 0.70, are the first structures to be formed in good folders (e.g. $q_{good}$ is 0.34 and 0.44 at state (IV) for them), while the {\em n-src} loop and the {\em Diverging turn} (with $q^{frozen}_{bad}$ respectively 0.25 and 0.20), the helix and the sheet $\beta_{1-5}$ are formed at lower temperatures in good folders.

\section{Discussion}

The thermodynamics of the SH3 domain has been widely studied by means of G\=o models \cite{go}, where only native contacts interact favourably. Conformational samplings of a $C_\alpha$ model where each native contact contribute to the total energy with the same energy $B_0=-1$ show a plot of the specific heat with a single, sharp peak at $0.63\cdot |B_0|$ \cite{kolya02}. A modified G\=o model where each pair of residues building a native contact interacts with a pair--dependent energy (the average being $B_0= -0.29$ and the standard deviation $\sigma_B= 0.37$) displays again essentially a single peak in $C_p$ centered at $0.85|B_0|$ (cf. ref. \cite{ludo}). The shape of the specific heat in the present model, where also non-native contacts are considered, is different and much more structured. First, there are a number of peaks which indicate that the G\=o interaction oversimplifies the thermodynamics of the chain. Although most of these peaks lie at non-biological temperature, this discrepancy makes one suspect that also other features of the thermodynamical states of the model protein can be oversimplified.

\begin{figure}
\centerline{\psfig{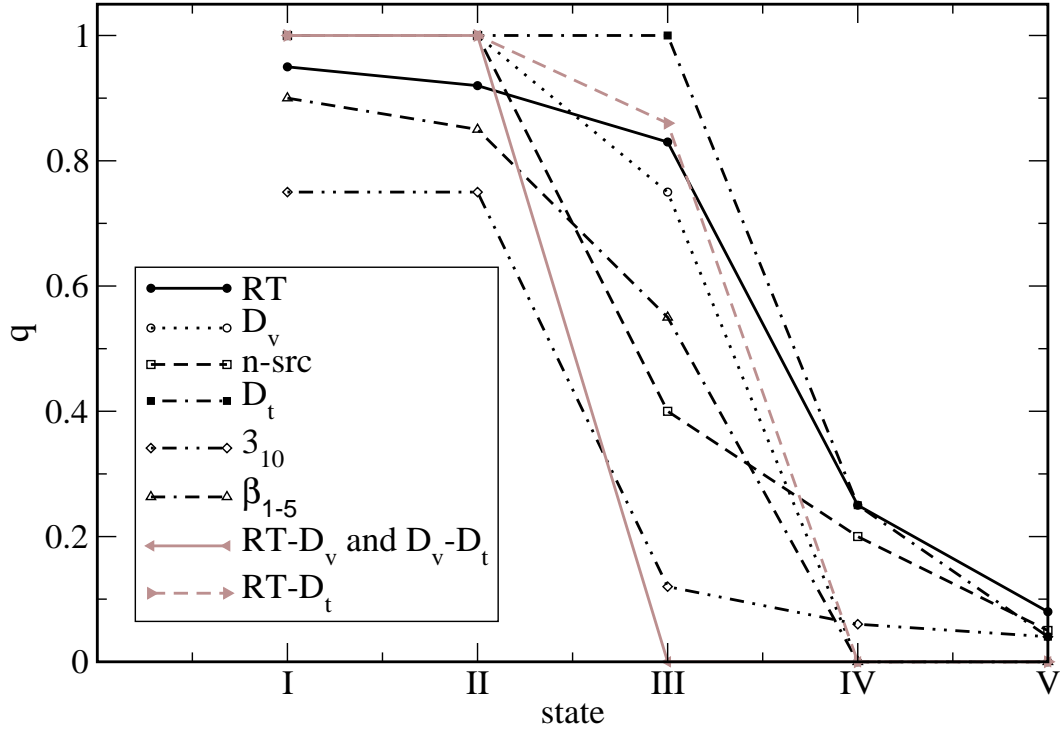}}
\caption{The structure content $q$ of the motifs of SH3 and between some of them (see Tables \protect\ref{tab_TS_G} and \protect\ref{table_new} of sequence $s_1$ in the different thermodynamical states.}
\label{fig_new}
\end{figure}

\begin{figure}
\centerline{\psfig{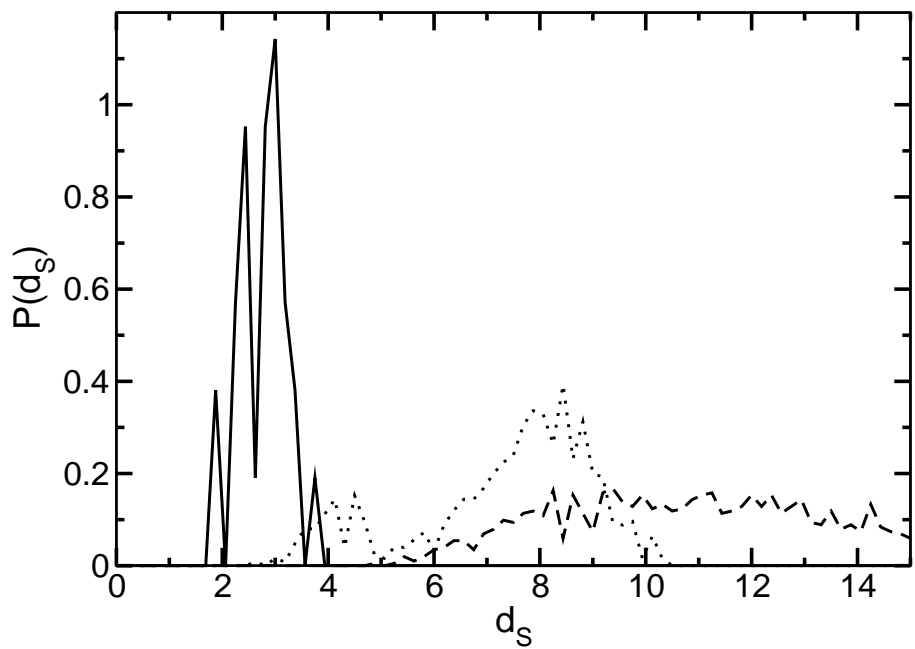}}
\caption{The distribution of dRMSD for pairs of conformations ($d_S$) associated with sequence s$_1$ at $T=0.10$ (solid curve), $T=0.20$ (dotted curve) and $T=0.50$ (dashed curve). }
\label{p_drmsd}
\end{figure}

\begin{figure}
\centerline{\psfig{file=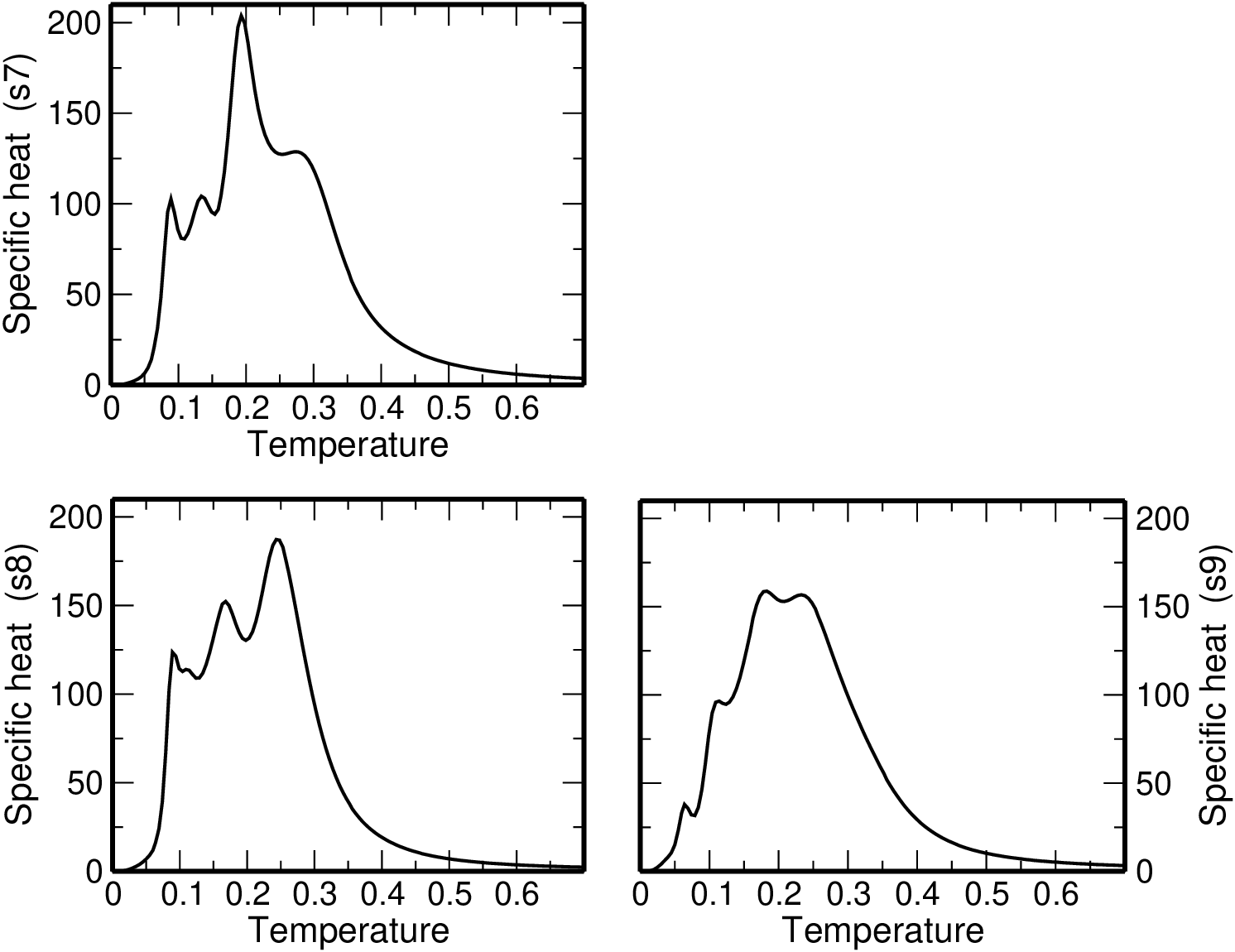,width=12cm}}
\caption{The specific heat $C_p(T)$ of the three random folders; sequence $s_7$ (top), sequence $s_8$ (bottom right) and sequence $s_9$ (bottom left).}
\label{calRand}
\end{figure}

\begin{figure}
\centerline{\psfig{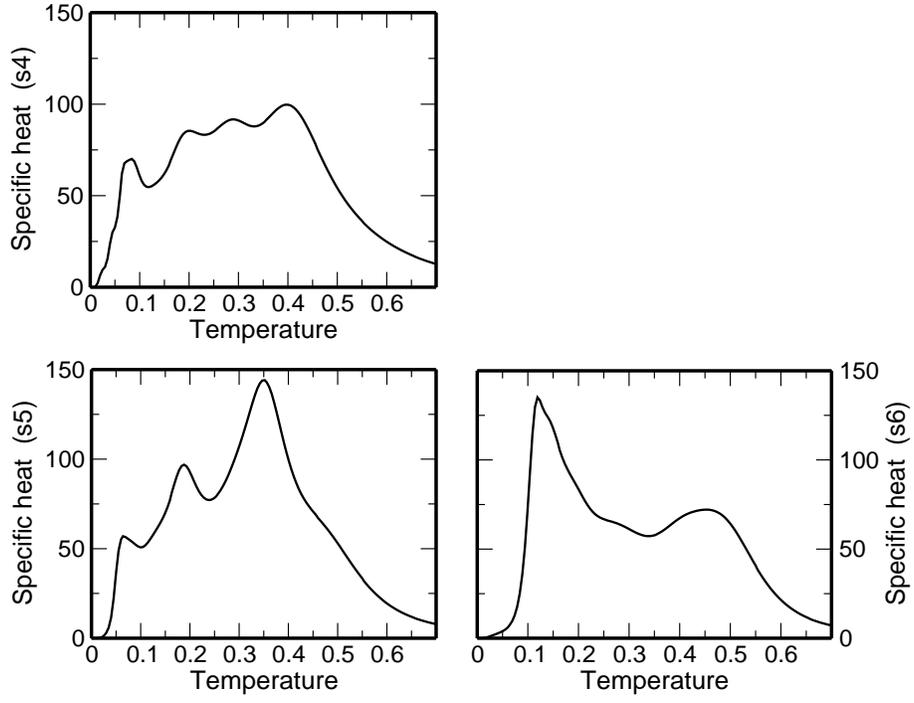}}
\caption{The specific heat $C_p(T)$ of the three bad folders; sequence $s_4$ (top), sequence $s_5$ (bottom right) and sequence $s_6$ (bottom left).}
\label{calBad}
\end{figure}

\begin{figure}
\centerline{\psfig{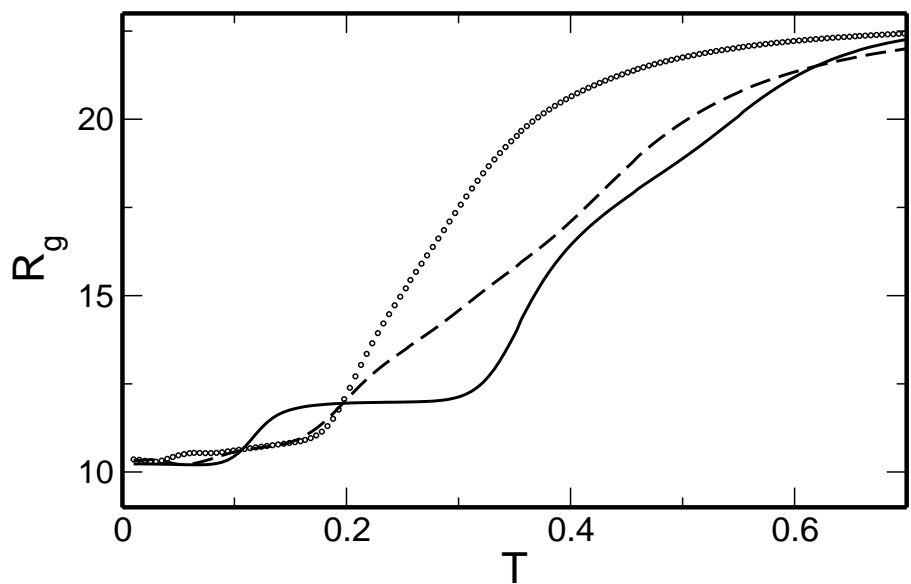}}
\caption{Radius of gyration as function of temperature for three sequences rapresentative of the three groups; a good folder (sequence $s_1$, continous solid curve), a bad folder (sequence $s_4$, dashed curve) and a random sequence (sequence $s_7$, open dots).}
\label{gyradCompa}
\end{figure}

\begin{figure}
\centerline{\psfig{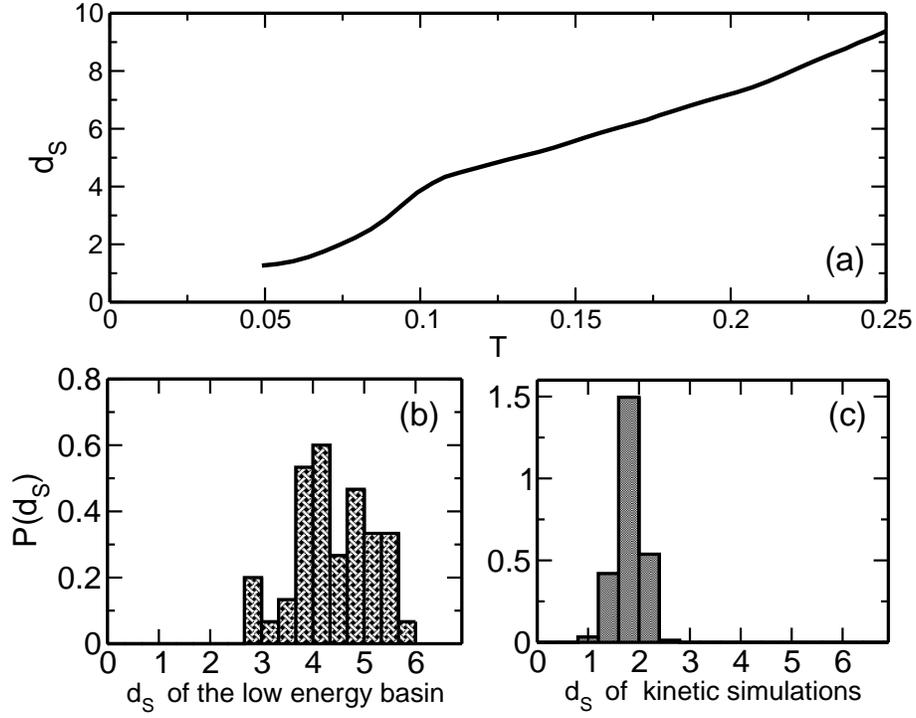}}
\caption{(a): $d_S(T)$ plot for $s_7$ (representative of random sequences) highlights the freezing transition at temperature $\approx 0.1$. (b): Distribution of $d_S$ for a set of low energy ($E < -39$) conformations for $s_7$. (c): Distribution of $d_S$ for states sampled in a low temperature ($T = 0.07$) kinetic simulation starting from one of these low energy states of sequence $s_7$.}
\label{dS_rand}
\end{figure}

\begin{figure}
\centerline{\psfig{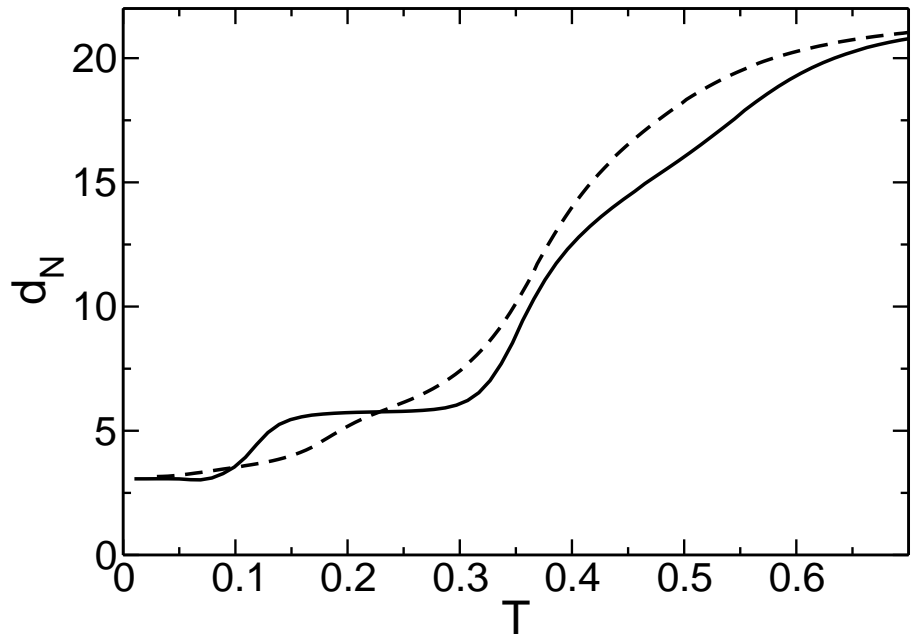}}
\caption{Variation with temperature of the dRMSD calculated with respect to the ground state structure ($d_N$) for a good folder (sequence $s_1$, continuous curve) and a bad folder (sequence $s_5$, dashed curve).}
\label{dGScompa}
\end{figure}

An interesting feature of the present model is that the biologically relevant unfolded state of the protein (state (III) of Fig. \ref{s1}, see also Fig. \ref{fig_new} and Tables \ref{tab_TS_G} and \ref{table_new}) is quite different from a random coil. First, it is rather compact, the average radius of gyration being $12\AA$, some $20\%$ larger than the native state. Note that the unfolded state predicted by our model is more compact and much more structured that that given by standard G\=o models, which have a $\bar{R}_g\simeq 25\AA$ and a total number of contacts which is approximately one fourth of that in the native state \cite{kolya02}. Second, a number of native and non-native contacts are rather stable in the unfolded state. In particular, the RT loop, the distal loop and the diverging turn result consistently populated. 

These results are in agreement with the NMR experiments of $\alpha$--spectrin SH3 under acidic conditions, which populate a denatured state \cite{kortemme}. This state displays NOE (Nuclear Overhauser Effect) signals in the region of the distal hairpin and of the preceding strand. Moreover, NMR studies of the drkN SH3 domain, an unstable protein which populate the unfolded state under non-denaturing conditions, indicate an even larger abundance of interactions \cite{mok} than the $\alpha$--spectrin experiments, involving the whole regions 9--20 and 25--48. The associated radius of gyration results of the order of $11\AA$. The radius of gyration resulting by the implementation of the NOEs is $\simeq 11\AA$, in agreement with the results of our model. Moreover, the 75 \% of the long range NOEs observed is non-native, a fact which highlights the necessity of accounting for non-native interactions in any model which aims at describing the unfolded state of a protein.

The complicated shape of the specific heat of good folders reveals a hierarchy of energy scales which can be useful to understand the folding of SH3. Some regions of the protein, such as the distal hairpin and the RT--loop, are structured even at very high temperature ($q=0.25$ for both structures at $T\approx 0.5$ (IV), cf. Table \ref{tab_TS_G} and Figs. \ref{s1} and \ref{fig_new}), indicating a remarkable propensity to fold independently on the rest of the protein. Using the language of \cite{wolynes}, we can see these regions as {\em foldons}.

Following ref. \cite{jchemphys3}, one can interpret these sequence of energy scales from a kinetic point of view, identifying high--temperature states as high--energy conformations at the beginning of the folding dynamics, and low--temperature states as the ending point of the dynamics. From this point of view, the regions of high--temperature conformations displaying native interactions can be regarded as the local elementary structures (LES) \cite{jchemphys2} which drive the folding kinetics. Note that this interpretation assign to non-native interactions an important role in the folding kinetics, as testified by the fact that their elimination in G\=o models affects the whole hierarchy of energy scales.
Within the framework of the hierarchical folding mechanism, the {\it RT--loop} and the distal hairpin act as (closed) LES in the language of ref. \cite{jpcm}. Their docking, taking place at the transition between III and II gives rise to the (post--critical) folding nucleus (FN), that is the minimum set of native contacts which brings the system over the highest barrier of free energy associated with the folding process.

Our results also show, complementing the findings of our previous work \cite{andrea}, that sequences obtained by minimization of the interaction energy at fixed native conformation not only fold fast but they also display realistic thermodynamical features. 

Bad sequences, although not being able to fold, display a consistent degree of structure in the same regions of the protein which in good sequences are ordered already at high temperature. This is consistent with the results obtained by means of lattice models \cite{jchemphys2}, where it is seen that contacts within and across local elementary structures are stabilized even in sequences obtained at low evolutionary pressure. In other words, such bad sequences have some of the kinetic features typical of good folders, but their energy is not low enough, so that they have to compete with a sea of alternative conformations.

When comparing our results with the random energy model (REM), we find some interesting differences. First of all our model shows a clear folding transition from a non-random globular state, which the REM assumes to require higher level of sidechain detail \cite{pande}. 
On the other hand we do not see any transition from a random globule to a folded globule for any sequence; sequences either achieve a specific globular configuration, from which they always fold into the native structure, or get trapped into a random globular state. 
Moreover, the coil--globule transition of our good folders corresponds to a peak in the specific heat, as expected in the case of first order transitions, in agreement with the non-random heteropolymer theory.
Our system contradicts also the prediction of the freezing temperature made by REM ($T^{theor}_{freeze}$ results to be much higher than the actual $T_{freeze}$) and the theoretical equivalence  $T^{cr}_{design}=T_{freeze}$, thus questioning the applicability of the self-averaging principle.

\begin{figure}
\centerline{\psfig{file=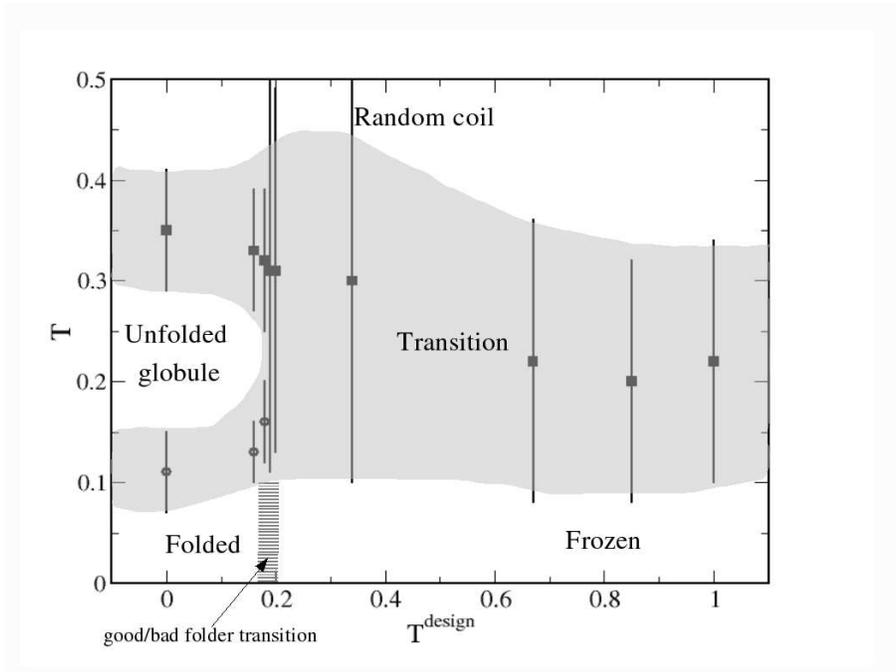,width=12cm}}
\caption{Phase diagram of the system. The behaviour in the conformational space as function of the design temperature is shown, where the two most relevant transitions are taken into account. The squares identify the centroids of the transitions (top of the $C_p$ peaks for the good folders), while the vertical lines span all over the transition regions for the 9 sequences studied. The shaded area marks the expected transition region at any value of the design temperature.}
\label{TofTdes}
\end{figure}

Our simulations provides information on the relationship between design temperature ($T_{design}$) and the thermodynamical behaviour of the corresponding sequences in our system. The outcome is summarized in Fig. \ref{TofTdes} \footnote{In that figure we have neglected the first ($(V) \rightarrow (IV)$)
and the last ($(II) \rightarrow (I)$) transitions for the good sequences and focused on the two most relevant ($(IV) \rightarrow (III)$ and $(III) \rightarrow (II)$)}. 
At high temperatures, sequences designed at any temperature adopt an highly disordered coil configuration. 
Decreasing the temperature, the behaviour becomes sequence--dependent. Good sequences first undergo a coil--globule transition to a partially ordered unfolded globule, and from there to the native state, while poorly designed sequences undergo a smooth transition to a frozen set of more or less disordered compact states ($R_g$ of the ground state is $\simeq 10 \AA$ for all the nine sequences).
The freezing/folding temperature remains roughly constant all over the range of  $T_{design}$, while the compaction begins at lower temperatures for higher $T_{design}$, in such a way that the coil -- globule and the globule -- frozen transitions ``merge'' into a wide coil -- frozen-globule transition region. 

This phase diagram can be compared with that arising from the theory of random heteropolymers (see Fig. 1 of ref. \cite{pande}). In overall agreement with these results, we also observe a random coil, a folded and a frozen phase. Furthermore, we also have found that the boundary between folded and glassy phases is a vertical line in the phase diagram and the freezing temperature essentially independent on $T_{design}$.
                                                                                                        
On the other hand, our phase diagram also shows important differences with the phase diagram describing the behaviour of random heteropolymers. Designed sequences (low $T_{design}$)  display  an unfolded globular phase showing most of the properties of the unfolded state measured in experiments. Furthermore,the results shown in Fig. \ref{TofTdes} display wide fluctuations between the states, the transition regions occupying most of the phase diagram. This feature, which is absent in the theoretical diagram calculated in the thermodynamic limit, highlights the important fact that proteins are finite, small systems, and that one should be careful in applying the tools of heteropolymer theory to real proteins.

\section{Conclusions} 
                                                                                                                             
We have used the SH3 domain as a benchmark to test the thermodynamical features of a protein model in which the energy function is non-trivial. Unlike G\=o models, this energy function does not contain directly any information on the conformational ground state of the protein, but only through the low (minimized) energy of the sequence in the native conformation. Furthermore, it allows for non-native interactions. The result is a richer set of states than those predicted by G\=o models. In particular, the unfolded state of selected sequences is not completely disordered, but is a globule where some of the native contacts are already stabilized.This fact has important implications in the folding kinetics of the protein.  The overall picture provided by standard theory of random heteropolymers is verified by our simulations but, again, the model displays richer features, where new phases are found and where the transition regions play an important role as a consequence of finite--size effects.

\clearpage
\newpage

%fig1

%fig2
\clearpage

%fig3
\clearpage

%fig4
\clearpage

%fig5
\clearpage

%fig6
\clearpage

%fig7
\clearpage

%fig8
\clearpage

%fig9
\clearpage

%fig10
\clearpage

%fig11
\clearpage

\clearpage
\begin{table}
\begin{tabular}{|c|c|c|l|}
\hline
$label$ & $E_{targ}$ & $E_{gs}$ &       {\bf{\scriptsize N----$\beta_1$---RT-$loop$------Dv----$\beta_2$-----n-src-----$\beta_3$-----Dt-----$\beta_4$-----$3_{10}$----$\beta_5$---C}} \\ \hline
%                                             123456789'123456789=123456789;123456789#123456789*1234567890
$s_1   $   & -37.80  &-46.96    &      {\tiny GLLLLAANNWWVTRTDEEKKDYVSSSSDDTQTGGYNIEGLIFFRQVVPPEAHTYYSSSTT}\\
$s_2   $   & -35.53  &-45.03    &      {\tiny QQHAASSSDDSDVFTVPPLGNLTNYYGIITKTTWLLFEGGAYTRNVDEEESSTLSVKYRW}\\
$s_3   $   & -34.85  &-44.92    &      {\tiny GDSAAAHQPERWWTTSSSEEPIYEVLLNVTTTFTRDVDSSDKVGFNGLLLQGTIYYNSKY}\\ \hline

$s_4   $   & -34.30  &-44.52    &      {\tiny QWAAHEEEDYRNFGTSSSYQGPGINSSFKTGYTTVDSDSLATRVVVDLLLILWEPKNYTT}\\
$s_5   $   & -33.65  &-45.02    &      {\tiny SGLNLEEPGKKYFRRTAAWFVEGSDSSVGTTTTNQHQTALLLWVSDDYYYIIVEPDSSTN}\\
$s_6   $   & -23.67  &-42.28    &      {\tiny DSSSSEERDIFYTTTWYYQQGPLNSLLLGTVKTVDDIYSSAKTRWVGAAHGPTEEFNLVN}\\ \hline

$s_7   $   &  -4.52  &-42.36    &      {\tiny NLILYEKLDNRFNKWWFLADSSPASGQVDRTTSTVSSTQEHTTYEEYVSGLGTIPDAVGY}\\
$s_8   $   &  +5.36  &-38.72    &      {\tiny LWYSSLEGGRVNLDTSSKVTPILSFAQGTDRVDDQEYTGIYWTVTEHTAEKYFNPNSALS}\\
$s_9   $   &  +8.26  &-40.92    &      {\tiny EYLSVIKTEDPKQSEYPSWLSEFFLLTIATGNTLYYDGVHAVTSSRNSGGDAVRNDTTWQ}\\
\hline
\end{tabular}
\caption{Sequences with selected energies $E_{targ}$ on the SH3 target conformation displayed in Fig. \protect\ref{native}. $E_{gs}$ is the energy of the ground state of the sequence. Sequences ($s_1$, $s_2$, $s_3$), ($s_4$, $s_5$, $s_6$) and ($s_7$, $s_8$, $s_9$) are good, bad and randomly generated folders, respectively.}
\label{tab_seq}
\end{table}

\begin{table}
\begin{tabular}{|l|c|c|c||c|c|c|c|c|c|}
\hline \hline
$label$ & $\bar{E}$ & $\bar{d}_N$ & $\bar{R_g}$ & $q(RT)$ & $q(D_v)$ & $q($n-src$)$ & $q(D_t)$ & $q(3_{10})$ & $q(\beta_{1-5})$ \\ \hline \hline
\multicolumn{10}{|l|}{State V: {\em ``coil''}}\\\hline
$s_1  $ & -2 &  21.0   & 22.5    &  0.08   & 0.00    & 0.05        & 0.04    &  0.04      & 0.00 \\
$s_2  $ & -2 &  20.9   & 22.4    &  0.05   & 0.03    & 0.05        & 0.03    &  0.02      & 0.00 \\
$s_3  $ & -2 &  21.3   & 22.7    &  0.05   & 0.03    & 0.03        & 0.05    &  0.00      & 0.00 \\ \hline
\hline
\multicolumn{10}{|l|}{State IV: {\em ``embryo''}}\\\hline
$s_1  $ & -10 &  14.8   & 18.2    &  0.25   & 0.00   & 0.20        & 0.25    &  0.06      & 0.00 \\
$s_2  $ & -13 &  14.4   & 17.6    &  0.45   & 0.00   & 0.00        & 0.50    &  0.15      & 0.00 \\
$s_3  $ & -14 &  13.0   & 16.9    &  0.33   & 0.17   & 0.22        & 0.56    &  0.00      & 0.00 \\ \hline
\hline
\multicolumn{10}{|l|}{State III: {\em ``globule''}}\\\hline
$s_1  $ & -37 &  5.7   & 12.0    &  0.83   & 0.75    & 0.40        & 1.00    &  0.12      & 0.55 \\
$s_2  $ & -35 &  5.9   & 12.0    &  0.85   & 0.66    & 0.20        & 1.00    &  0.15      & 0.50 \\
$s_3  $ & -32 &  6.2   & 12.2    &  0.80   & 0.70    & 0.13        & 0.93    &  0.42      & 0.40 \\ \hline
\hline
\multicolumn{10}{|l|}{State II: {\em ``folded''}}\\\hline
$s_1  $ & -45 &  3.0   & 10.2    &  0.92   & 1.00    & 1.00        & 1.00    &  0.75      & 0.85 \\
$s_2  $ & -42 &  3.6   & 10.2    &  0.85   & 0.90    & 0.70        & 1.00    &  0.45      & 0.78 \\
$s_3  $ & -41 &  3.9   & 10.3    &  0.95   & 1.00    & 0.85        & 1.00    &  0.50      & 0.70 \\ \hline 
\hline
\multicolumn{10}{|l|}{State I: {\em ``frozen''}}\\\hline
$s_1  $ & -47 &  3.0   & 10.2    &  0.95   & 1.00    & 1.00    & 1.00    &  0.75      & 0.90 \\
$s_2  $ & -45 &  3.6   & 10.0    &  0.95   & 0.92    & 0.88    & 1.00    &  0.66      & 0.88 \\
$s_3  $ & -45 &  4.0   & 10.1    &  1.00   & 1.00    & 0.85    & 1.00    &  0.45      & 0.70 \\ \hline
\hline
\end{tabular}
\caption{The average energies and structural features are here summarized for the three good folders at the five thermodynamically relevant states. From left to right, columns show the average value of the energy, of the dRMSD from the native state ($d_N$), of the radius of gyration ($R_g$) and the structure content $q$ of  six secondary structures of SH3 (that is, the {\em RT-loop} ($RT$), the {\em Distal hairpin} ($D_t$), the {\em Diverging turn} ($D_v$), the {\em n-src} loop (n-src), the helix $3_{10}$ and the sheet $\beta_{1-5}$).}
\label{tab_TS_G}
\end{table}

\begin{table}
\begin{tabular}{|l|c|c|c|c|c|c|}\hline
	& $\beta_{1-5}$		& $RT$		& $D_v$		& n-src		& $D_t$		& $3_{10}$ \\\hline
\multicolumn{7}{|l|}{{\em Frozen} state (I)}\\\hline
$\beta_{1-5}$ & 		& -		& 0.86		& 1		& -		& -	    \\
$RT$	& 			&		& 1		& -		& 1		& -	    \\
$D_v$	&  			&		&		& 1		& 1		& -	    \\
n-src	&			&		&		&		& 1		& 1	    \\
$D_t$	&			&		&		&		&		& -	    \\\hline
\multicolumn{7}{|l|}{{\em Native} state (II)}\\\hline
$\beta_{1-5}$ & 		& -		& 0.79		& 0.5		& -		& -	    \\
$RT$	& 			&		& 1		& -		& 1		& -	    \\
$D_v$	&  			&		&		& 0		& 1		& -	    \\
n-src	&			&		&		&		& 1		& 1	    \\
$D_t$	&			&		&		&		&		& -	    \\\hline
\multicolumn{7}{|l|}{{\em Unfolded globule} state (III)}\\\hline
$\beta_{1-5}$ & 		& -		& 0.71		& 0.5		& -		& -	    \\
$RT$	& 			&		& 0		& -		& 0.86		& -	    \\
$D_v$	&  			&		&		& 0		& 0		& -	    \\
n-src	&			&		&		&		& 1		& 1	    \\
$D_t$	&			&		&		&		&		& -	    \\\hline
\multicolumn{7}{|l|}{{\em Embryo globule} state (IV)}\\\hline
$\beta_{1-5}$ & 		& -		& 0.56		& 0		& -		& -	    \\
$RT$	& 			&		& 0		& -		& 0		& -	    \\
$D_v$	&  			&		&		& 0		& 0		& -	    \\
n-src	&			&		&		&		& 0		& 0	    \\
$D_t$	&			&		&		&		&		& -	    \\\hline
\end{tabular}
\caption{The structure content $q$ associated with contacts between SH3--structures of sequence $s_1$ are reported for the different thermodynamical states.}
\label{table_new}
\end{table}

\begin{table}
\begin{tabular}{|l|c|c|c||c|c|c|c|c|c|}
\hline \hline
$label$ & $\bar{E}$ & $\bar{d}_N$ & $\bar{R_g}$ & $q(RT)$ & $q(D_v)$ & $q($n-src$)$ & $q(D_t)$ & $q(3_{10})$ & $q(\beta_{1-5})$ \\ \hline \hline
\multicolumn{10}{|l|}{{\em ``Coil''} state}\\\hline
$s_7  $ & -2 &  21.2   & 22.3       &  0.05   & 0.02    & 0.00       & 0.00    &  0.06      & 0.00 \\
$s_8  $ & -1 &  21.9   & 22.6       &  0.02   & 0.00    & 0.07       & 0.00    &  0.00      & 0.00 \\
$s_9  $ & -2 &  21.5   & 22.6       &  0.03   & 0.08    & 0.02       & 0.03    &  0.02      & 0.00 \\ \hline
\hline
\multicolumn{10}{|l|}{{\em ``Globule''} state}\\\hline
$s_7  $ & -37 &  5.6   & 10.5       &  0.32   & 0.15    & 0.45       & 0.38    &  0.15      & 0.11 \\
$s_8  $ & -35 &  6.3   & 10.0       &  0.15   & 0.22    & 0.12       & 0.20    &  0.10      & 0.06 \\
$s_9  $ & -34 &  6.0   &  9.6       &  0.40   & 0.27    & 0.53       & 0.07    &  0.08      & 0.05 \\ \hline 
\hline
\multicolumn{10}{|l|}{{\em ``Frozen''} state}\\\hline
$s_7  $ & -42 &  5.4   & 10.3       &  0.14   & 0.22    & 0.20       & 0.45    &  0.08      & 0.14 \\
$s_8  $ & -39 &  5.9   &  9.8       &  0.22   & 0.25    & 0.05       & 0.30    &  0.03      & 0.05 \\
$s_9  $ & -41 &  5.9   &  9.3       &  0.44   & 0.38    & 0.72       & 0.03    &  0.12      & 0.08 \\ \hline
\hline
\end{tabular}
\caption{The average energies and structural features are displayed for the three random sequences at the three thermodynamically relevant states. From left to right, the different columns show the average value of the energy, of the dRMSD from the native state ($d_N$), of the radius of gyration ($R_g$) and the structural content $q$ of six secondary structures of SH3 (namely, the {\em RT-loop} ($RT$), the {\em Distal hairpin} ($D_t$), the {\em Diverging turn} ($D_v$), the {\em n-src} loop (n-src), the helix $3_{10}$ and the sheet $\beta_{1-5}$).}
\label{tab_TS_R}
\end{table}

\begin{table}
\begin{tabular}{|l|c|c|c||c|c|c|c|c|c|}
\hline \hline
$label$ & $\bar{E}$ & $\bar{d}_N$ & $\bar{R_g}$ & $q(RT)$ & $q(D_v)$ & $q($n-src$)$ & $q(D_t)$ & $q(3_{10})$ & $q(\beta_{1-5})$ \\ \hline \hline
\multicolumn{10}{|l|}{{\em ``Coil''} state}\\\hline
$s_4  $ & -2 &  20.9   & 22.3       &  0.10   & 0.05    & 0.00        & 0.12    &  0.00      & 0.00 \\
$s_5  $ & -1 &  21.3   & 22.6       &  0.07   & 0.05    & 0.05        & 0.15    &  0.00      & 0.00 \\
$s_6  $ & -1 &  21.3   & 22.6       &  0.08   & 0.03    & 0.03        & 0.10    &  0.00      & 0.00 \\ \hline
\hline
\multicolumn{10}{|l|}{{\em ``Globule''} state}\\\hline
$s_4  $ & -36 &  4.2   & 10.9       &  0.70   & 0.55    & 0.45        & 0.77    &  0.00      & 0.04 \\
$s_5  $ & -37 &  5.4   & 11.2       &  0.75   & 0.35    & 0.10        & 0.82    &  0.08      & 0.07 \\
$s_6  $ & -33 &  6.2   & 10.7       &  0.45   & 0.14    & 0.33        & 0.56    &  0.12      & 0.10 \\ \hline 
\hline
\multicolumn{10}{|l|}{{\em ``Frozen''} state}\\\hline
$s_4  $ & -45 &  4.5   & 10.3       &  0.67   & 0.25    & 0.20        & 0.80    &  0.07      & 0.04 \\
$s_5  $ & -45 &  5.2   & 10.6       &  0.70   & 0.30    & 0.10        & 0.80    &  0.12      & 0.15 \\
$s_6  $ & -42 &  6.0   & 10.6       &  0.40   & 0.20    & 0.30        & 0.50    &  0.15      & 0.18 \\ \hline
\hline
\end{tabular}
\caption{The average energies and structural features are here summarized for the three bad folders at the three thermodynamically relevant states. From left to right, columns show the average value of the energy, of the dRMSD from the native state ($d_N$), of the radius of gyration ($R_g$) and the structural content $q$ of six secondary structures of SH3 (namely, the {\em RT-loop} ($RT$), the {\em Distal loop} ($D_t$), the {\em Diverging turn} ($D_v$), the {\em n-src} loop (n-src), the helix $3_{10}$ and the sheet $\beta_{1-5}$).}
\label{tab_TS_B}
\end{table}

\end{document}